\pdfoutput=1
\documentclass[aps,amsmath,preprint,superscriptaddress,nofootinbib]{revtex4-1}
\usepackage{geometry}
\usepackage[utf8]{inputenc}

\usepackage[T1]{fontenc}
\usepackage{lmodern}
\usepackage{verbatim}
\usepackage{amsmath,amssymb,amsfonts}	
\usepackage{graphicx}
\usepackage{color}
\usepackage{xcolor}
\usepackage[colorlinks=true,citecolor=blue,linkcolor=purple,urlcolor=purple]{hyperref}
\usepackage{booktabs}
\usepackage{graphicx}
\usepackage{appendix}
\usepackage{subcaption}
\usepackage{slashed}
\newcommand{\red}[1]{{\color{red} #1}}

\begin{document}
\preprint{ULB-TH/22-15}

\title{A closer look at the seesaw-dark matter correspondence}
\author{Rupert Coy}
\affiliation{Service de Physique Théorique, Université Libre de Bruxelles,\\
Boulevard du Triomphe, CP225, 1050 Brussels, Belgium}
\author{Aritra Gupta}
\affiliation{Service de Physique Théorique, Université Libre de Bruxelles,\\
Boulevard du Triomphe, CP225, 1050 Brussels, Belgium}

\begin{abstract}
In an earlier work \cite{Coy:2021sse}, we showed that in very simple neutrino portal-like extensions of the Standard Model it is possible to achieve a one-to-one correspondence between dark matter physics and the seesaw parameters controlling the genesis of neutrino masses. 
Notably, this can occur both when the dark matter is produced via freeze-in and relativistic freeze-out. 
In this article, we carry out a detailed phenomenological study of such scenarios. Specifically, we find the allowed regions for the neutrino portal coupling within which the correspondence is valid. 
We further constrain the parameter space from various observational and theoretical considerations. 
Within this, we derive the distribution function of a doubly frozen-in particle in order to more accurately compute its free-streaming horizon. 
\end{abstract}

\maketitle
\tableofcontents

\section{Introduction}
The nature of dark matter (DM) and the origin of neutrino masses are two of the most important open problems in particle physics, and provide clear evidence for the necessity of physics beyond the Standard Model (SM). 
While a priori there is no reason why the resolution to these dual problems should have a common origin, it remains an attractive and economical possibility. 
The simplest way to link dark matter and neutrino masses is in models of keV-scale sterile neutrino dark matter \cite{Dodelson:1993je,Shi:1998km,Asaka:2005pn}, in which the type-I seesaw mechanism directly provides a DM candidate. 
However, the allowed parameter space for such a simple case has been very constrained by a combination of x-ray and Lyman-$\alpha$ constraints \cite{Boyarsky:2008xj,Drewes:2016upu,Irsic:2017ixq} (see e.g. \cite{Shaposhnikov:2006xi,Kusenko:2006rh,Merle:2013gea,Merle:2013wta,Frigerio:2014ifa,Biswas:2016bfo,Lucente:2021har,Datta:2021elq,Coy:2022unt} for a sample of models which overcome this in different ways).

Perhaps the next most minimal way to connect neutrino masses and DM is via neutrino-portal DM models, where sterile neutrinos of the type-I seesaw mechanism act not as the DM itself, but as a mediator between the SM and the dark sector. 
The parameter space becomes significantly broader, since there are additional fields, and the DM itself is not confined to the keV range. 
In many cases, the sterile neutrino is effectively decoupled from the SM due to small portal couplings, while the DM may be neutral or have some SM or dark charge, see for instance \cite{Pospelov:2007mp,Falkowski:2009yz,Aoki:2015nza,Gonzalez-Macias:2016vxy,Becker:2018rve,Berlin:2018ztp,Bandyopadhyay:2020qpn,Chianese:2021toe}. 
In this paper we focus on neutrino-portal DM where the dark sector consists of an additional singlet fermion and scalar, both neutral with respect to the SM and coupled to the sterile neutrino via a single Yukawa interaction.

It was recently noted in \cite{Coy:2021sse} that this scenario is particularly simple and predictive when the DM particle, $\chi$, is lighter than the sterile neutrino, which is itself lighter than the electroweak scale, i.e. 
\begin{equation}
    m_\chi < m_N < m_W \, .
\end{equation}
Then, quite generically, the DM abundance depends only on the seesaw parameters and the DM mass itself, being essentially independent of the other dark sector parameters. 
In fact, as outlined in \cite{Coy:2021sse}, this occurs in two separate regimes. 
When the dark sector coupling is large, the DM abundance is determined by relativistic freeze-out, while when the dark sector coupling is very small, it is produced via sequential freeze-in. 
Nevertheless, in both cases the relic abundance turns out to be independent of this dark sector coupling. 
Moreover, such a model predicts a very light (active) neutrino mass, $m_{\nu 1} \ll$ meV, as well as the possibility of an observable neutrino line, a smoking-gun astrophysical DM signature.

In this paper, we follow up on Ref. \cite{Coy:2021sse} and perform a more thorough phenomenological analysis. 
We believe that this is warranted for a few reasons. 
Firstly, the attractiveness of the simple scenario justifies a broader and more thorough study than was presented in the original paper. 
Secondly, while general arguments were presented for the two regimes mentioned in the previous paragraph, a proper numerical analysis is required to find under exactly which conditions this holds true. 
Thirdly, since the neutrino portal is a popular model in its own right and a good representative of many freeze-in and freeze-out models involving a portal to the SM, some of our results\textemdash such as the distribution function of DM produced by sequential freeze-in\textemdash may be of general interest. 

In section \ref{sec:Model}, we briefly outline the neutrino portal model under consideration. 
Then in section \ref{sec:FO}, we turn to the relativistic freeze-out regime (which occurs when the dark sector interaction is sizeable), and analyse the various bounds and present our key results in Fig. \ref{fig:relFO}. 
Section \ref{sec:FI} is dedicated to the sequential freeze-in regime (which applies when the dark sector interacts feebly), and our results for this scenario are summarised in Fig. \ref{fig:FI}. 
Some technical details are discussed further in appendix \ref{sec:collisions}.

\section{Model}
\label{sec:Model}
We begin with the type-I seesaw mechanism with a single sterile neutrino, 
\begin{align}
    \mathcal{L} = \mathcal{L}_\text{SM} + i \overline{N_R}\slashed{\partial}N_R - \frac{1}{2} m_N (\overline{N_R} N_R^c + \overline{N_R^c} N_R) - (Y_\nu \overline{N_R} \tilde{H}^\dagger L + h.c.)  \, .
\end{align}
Here, $Y_\nu$ is a $1\times 3$ row vector, with entries $Y_{\nu i}$ for $i=e,\mu,\tau$. 
These are assumed to be very small, the $\mathcal{O}(10^{-13}-10^{-9})$ size required for successful freeze-in of DM \cite{Hall:2009bx}. 
This leads to a neutrino mass of
\begin{equation}
    m_{\nu_1} = \sum \limits_i \frac{Y_{\nu i}^2 v^2}{2m_N} \, .
\end{equation}
Given the smallness of $Y_{\nu i}$, the constraints discussed in the following sections, and the current neutrino mass splitting data \cite{Esteban:2020cvm}, this mass eigenstate must be the lightest neutrino, with $m_{\nu_1} \ll \text{eV}$. 
At least two more sterile neutrinos are required for a seesaw mechanism which correctly reproduces neutrino mass data, however we assume that these have much larger masses, $m_{N_2, N_3,\ldots} \gg m_N$, and are therefore decoupled.

In addition to the sterile neutrino, we add a real scalar, $\phi$, and Majorana fermion, $\chi$, both of which are singlets of the SM gauge group,
\begin{align}
    \mathcal{L}_{\rm dark} = i \overline{\chi} \slashed{\partial} \chi + \frac{1}{2} (\partial_\mu \phi)^2 - \frac{1}{2}m_\chi (\overline{\chi} \chi^c + \overline{\chi^c} \chi) - Y_\chi (\overline{N_R} \phi \chi + h.c.) - V(\phi) \, ,
\end{align}
where $V(\phi)$ is the $\phi$ potential. 
As discussed in \cite{Coy:2021sse}, these couplings can be justified by various possible global or gauge symmetries. 
In this model, both $\phi$ and $\chi$ are DM candidates, depending on their relative masses. 
Notably, in the relativistic freeze-out scenario (large values of $Y_\chi$, section \ref{sec:FO}), the lighter particle is the DM, while in the sequential freeze-in scenario (tiny values of $Y_\chi$, section \ref{sec:FI}), both are DM with the heavier one giving the dominant contribution.


\section{Relativistic freeze-out of DM}
\label{sec:FO}
First, we consider DM production via relativistic freeze-out. 
Initially, sterile neutrinos are mainly produced through decays of the SM gauge bosons, $Z \to N\nu$ and $W^\pm \to N \ell^\pm$, since $m_N < m_W$ \cite{Coy:2021sse}. 
The role of Higgs decays, and of scatterings such as $\ell^+ \ell^- \to N\nu$, are subdominant. 
Subsequently, decays and annihilations such as $N \leftrightarrow \chi \phi$, $NN \leftrightarrow \chi \chi$, $NN \leftrightarrow \phi \phi$ and $\chi \chi \leftrightarrow \phi \phi$ lead to the production of $\chi$ and $\phi$ particles. 
For sufficiently large values of $Y_\chi$, these processes equilibrate and thus the dark sector particles form a thermal bath with temperature $T' < T$. 
Here we assume that the fermion $\chi$ is DM, with $m_\chi < m_\phi$, however the situation would be very similar if the $\phi$ were considered the DM instead. 
In the limit that $m_N \gg m_\chi$, interactions involving the DM will drop out of equilibrium at some temperature $m_N \gg T'_{\rm dec} \gg m_\chi$ (see \ref{ssec:ydep} for the calculation of $T'_{\rm dec}$). 
This is analogous to the decoupling of neutrinos from the SM at a temperature $m_W \gg T_{\nu \text{ dec}} \gg m_\nu$, see \cite{Hambye:2020lvy,Coy:2021ann} for a more general discussion of the relativistic freeze-out scenario. 
Since in this case the DM freezes out while it has a relativistic number density, its relic abundance is given by the simple relation
\begin{equation}
    \Omega_\chi h^2 = 0.12 \, \frac{g_\chi m_\chi}{6 \text{ eV}} \left( \frac{g_{*s,0}}{g_{*s,\text{dec}}} \right) \xi_{\rm dec}^3 \, ,
    \label{eq:relicnonrel}
\end{equation}
where $g_\chi=2$ is the DM degrees of freedom, $g_{*s,0}$ and $g_{*s,\text{dec}}$ are the relativistic entropic degrees of freedom today and at the time of DM freeze-out, respectively, and $\xi_{\rm dec} = T'_{\rm dec}/T_{\rm dec}$ is the temperature ratio of the two sectors when they decouple.

The evolution of the dark sector temperature, $T'$, is determined by computing the energy injection into this sector, see for instance \cite{Chu:2011be,Coy:2022unt}. 
As stated above, the production of dark sector particles is dominated by SM gauge boson decays to sterile neutrinos, which gives
\begin{align}
    a^{-4} \frac{d(\rho' a^4)}{dt} &\simeq \sum \limits_{X=W,Z} \frac{g_X m_X^3 T}{2\pi^2} \Gamma(X \to N) K_2(m_X/T) \, ,
\end{align}
where $\rho' = \pi^2 g_{*,HS} T'^4/30$ is the dark sector energy, with $g_{*,HS} = 9/2$ when the $N$, $\chi$ and $\phi$ are all relativistic. 
This simple ODE can be solved numerically, and the ratio $\xi \equiv T'/T$ becomes approximately constant for $T \lesssim m_W$, after which most of the dark sector particles have decayed. 
The limiting value is\footnote{
Here and throughout the paper we neglect flavour, which is unimportant for our analysis.  
We therefore consistently write $|Y_\nu^2|$ in place of $\sum \limits_{i=e,\mu,\tau} |Y_{\nu i}|^2$ for convenience.} 
\begin{equation}
     \xi \simeq 0.014 \left(\frac{|Y_\nu|^2}{10^{-24}} \right)^{1/4} \sqrt{\frac{10 \text{ GeV}}{m_N}} \, .
     \label{eq:xiFin}
\end{equation}
Substituting this into Eq. \eqref{eq:relicnonrel}, we find that the correct abundance is achieved when 
\begin{equation}
    m_\chi \simeq 9 \, \frac{g_{*s,\text{dec}}}{g_{*s,0}} \left(\frac{10^{-24}}{|Y_\nu|^2}\right)^{3/4}  \left(\frac{m_N}{10 \text{ GeV}} \right)^{3/2} \text{MeV} \, .
    \label{eq:mchirelFO}
\end{equation}
Notably, there is a direct relationship between the DM mass and the seesaw parameters $Y_\nu$ and $m_N$. 
This result is independent of the coupling $Y_\chi$ and the $\phi$ mass, given the previous assumptions. 
Having outlined this scenario, we now find the region of parameter space for which this simple relationship holds and which survives various observational bounds.

\subsection{$Y_\chi$-independent constraints}
%

\begin{figure}[!t]
    \centering
    \includegraphics[width=0.49\textwidth]{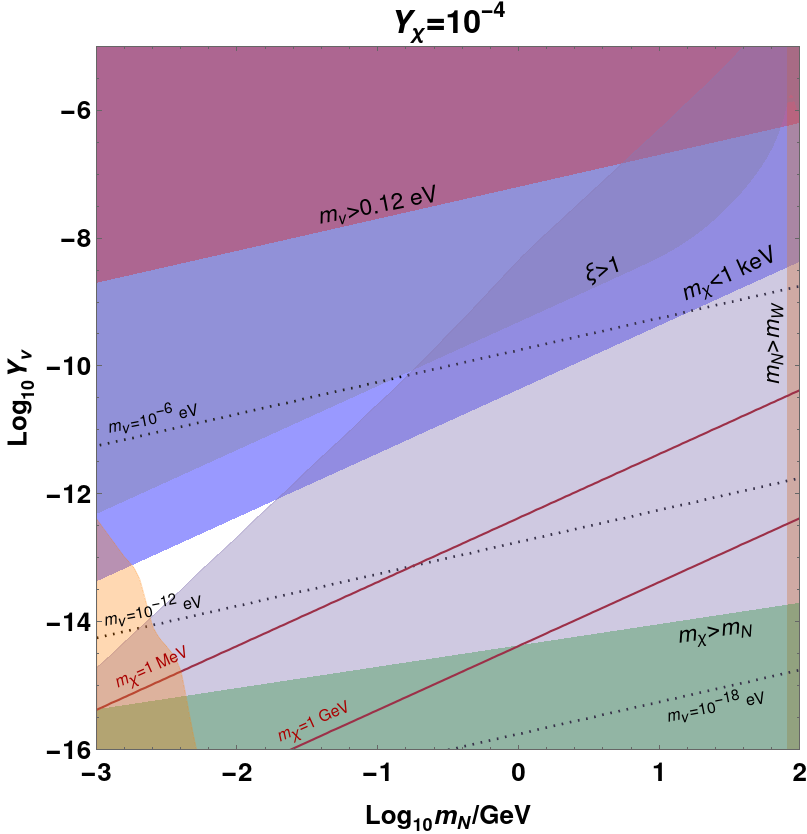}~
    \includegraphics[width=0.49\textwidth]{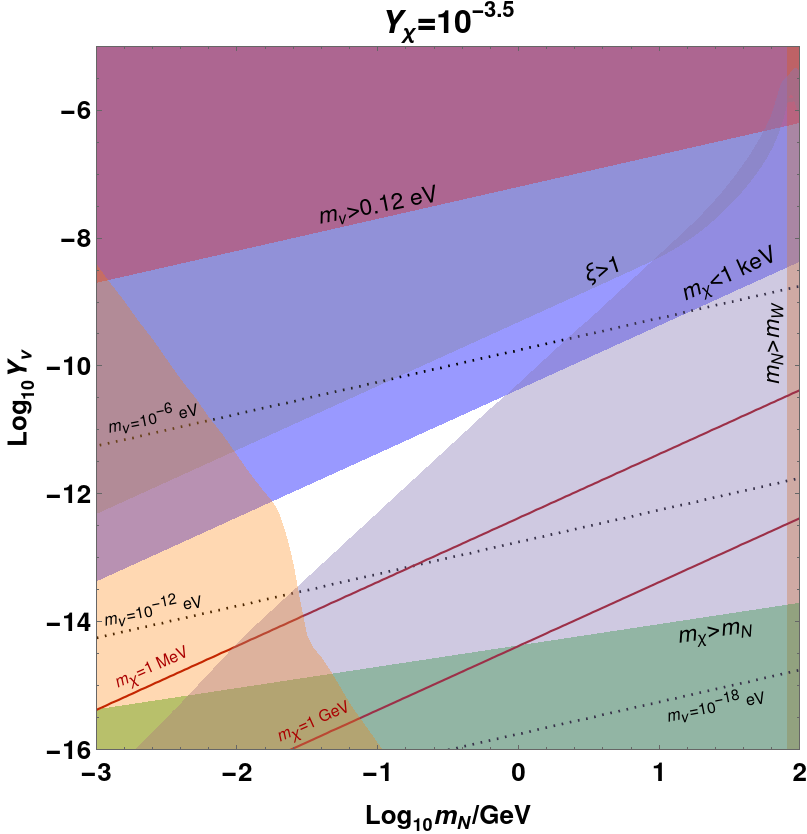}
    \includegraphics[width=0.49\textwidth]{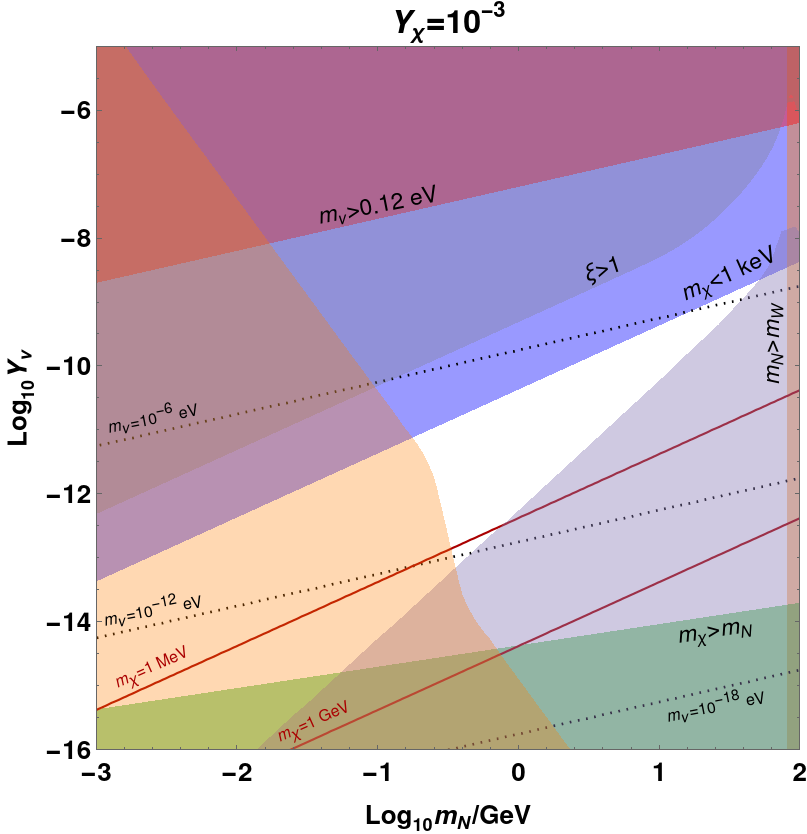}
    \includegraphics[width=0.49\textwidth]{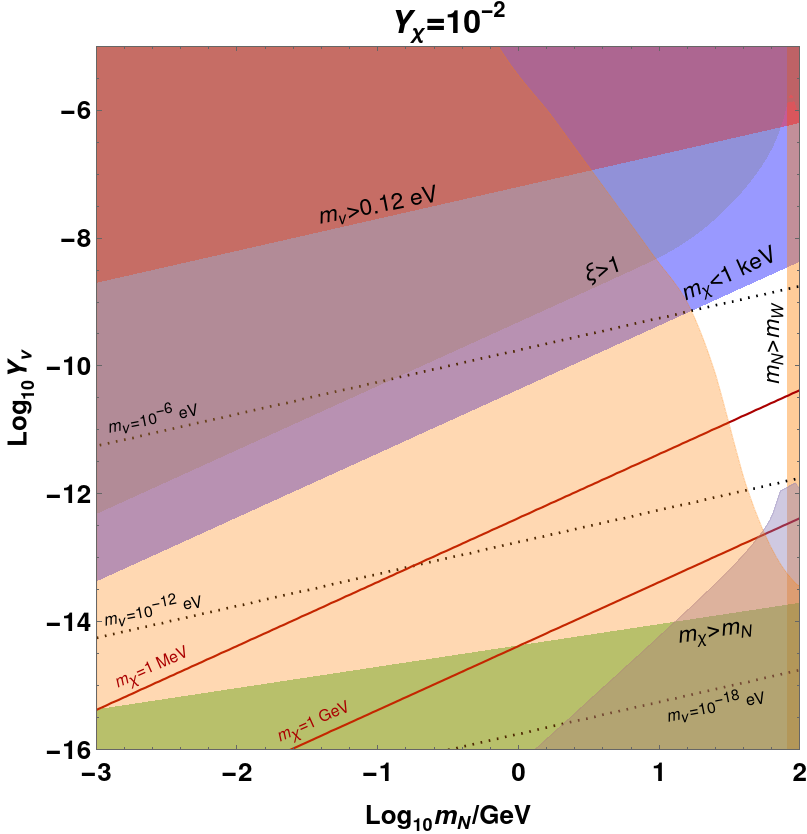}
    \caption{Allowed region for relativistic freeze-out scenario for different values of dark sector coupling $Y_\chi$. 
    The constraints and contours are indicated in the plots, except i) the light orange region from the bottom left which excludes parameter space where the DM does not freeze-out relativistically, and ii) the light purple region from the bottom right which excludes parameter space where dark sector thermalisation is never achieved. 
    }
    \label{fig:relFO}
\end{figure}

There are four relevant constraints which are independent of the dark sector coupling, $Y_\chi$. 
First of all, the assumed mass hierarchies must be respected. 
We enforce that $m_W > m_N > m_\chi$ in order to be consistent with the fact that the sterile neutrinos are produced by gauge boson decays and the DM is produced by sterile neutrino decays and annihilations. 
The regions excluded by this requirement are shown in Fig. \ref{fig:relFO} as the thin orange strip on the right edge and the green region at the bottom of each panel. 
Secondly, in order to satisfy the Tremaine-Gunn bound \cite{Tremaine:1979we}, we set $m_\chi \gtrsim 1$ keV (note, however, that if the scalar $\phi$ were the DM with $m_\phi < m_\chi$, then this bound would not apply). 
This constraint is given by the blue area in each panel, recalling the relation in Eq. \eqref{eq:mchirelFO}. 
It is similar to the one obtained from Lyman-alpha data by \cite{Irsic:2017ixq}. 
They found $m_{\rm DM} > 5.3$ keV at $2\sigma$ C.L., however the constraint becomes weaker proportionally to $T'/T$. 
Unlike the Tremaine-Gunn limit, this is insensitive to whether the DM is fermionic or bosonic. 
Note also that contours of $m_\chi = 1$ MeV and 1 GeV are displayed in red in Fig. \ref{fig:relFO}. 
In all the allowed (white) regions, the DM is sub-GeV, and for $Y_\chi \lesssim 10^{-3}$, it is even sub-MeV.

A third, relatively weaker bound is acquired by enforcing that the neutrino mass cannot be too large. 
As discussed previously, the lightest neutrino mass will be no larger than $|Y_\nu|^2 v^2/(2m_N)$. 
We impose that this value is not larger than the limit obtained by Planck plus BAO data on the sum of the neutrino masses, $\Sigma m_\nu < 0.12$ eV at $95\%$ C.L. \cite{Planck:2018vyg}. 
The region of parameter space which gives too large a value of $m_\nu$ is shaded in purple. 
The dotted black lines correspond to the contours of $m_{\nu 1} = 10^{-6,12,18}$ eV, and thus we see that the allowed regions permit $10^{-13}$ eV $\lesssim m_{\nu 1} \lesssim 10^{-5}$ eV. 
Thus, the model unambiguously predicts a very tiny lightest neutrino mass. 
While a discovery of such a light neutrino does not seem possible in the near future, this scenario is falsifiable in experiments which probe the absolute neutrino mass scale, for instance at KATRIN \cite{KATRIN:2019yun}.

An additional constraint comes from considering the thermalisation process. 
We insist that the gauge boson decays to sterile neutrinos do not thermalise, so that the SM and dark sectors do not come into equilibrium, which would lead to $\xi \to 1$. 
This would complicate the DM freeze-out and also lead to an unacceptably large contribution to $N_{\rm eff}$ at BBN compared to observations if $m_\chi \lesssim$ 5 MeV \cite{Fields:2019pfx}. 
From Eq. \eqref{eq:xiFin}, this corresponds to a contour of constant $Y_\nu/m_N$ that is independent of $Y_\chi$, except when $m_N \simeq m_W$ and this estimate for $\xi$ breaks down. 
The excluded region is shown in grey in Fig. \ref{fig:relFO}.

\subsection{$Y_\chi$-dependent constraints}
\label{ssec:ydep}
All the bounds discussed so far have been independent of $Y_\chi$ and thus rule out the same regions in the four panels of Fig. \ref{fig:relFO}. 
We now turn to two key $Y_\chi$-dependent constraints. 
The first comes from the consistency condition that the $\chi$ does indeed freeze-out relativistically. 
The last process to go out of equilibrium is $\chi \chi \leftrightarrow \phi \phi$, since scatterings and annihilations involving the $N$ as an external state become Boltzmann-suppressed at $T' < m_N$, while the $N$-mediated $\chi \chi \leftrightarrow \phi \phi$ has a milder suppression by factors of $T'/m_N$. 
The $\chi$ freezes out relativistically, i.e. while it has number density $n_\chi \sim T'^3$, if this process goes out of equilibrium when $T'_{\rm dec} > m_\chi$. 
The scattering cross-section is
\begin{equation}
    \sigma(\chi \overline{\chi} \to \phi \phi) \simeq \frac{Y_\chi^4}{
 16 \pi s^2} \left( \frac{6 m_N^4 + 8 m_N^2 s + s^2}{2 m_N^2 + s} \log(1 + \frac{s}{m_N^2}) - \frac{ s (3 m_N^2 + 2 s)}{m_N^2 + s} \right)
\end{equation}
neglecting $m_\chi, m_\phi \ll m_N$. 
The thermally-averaged rate can be computed via the standard formalism \cite{Gondolo:1990dk}. 
For numerical ease as we scan over a large number of points, we utilise a fit function, writing
\begin{equation}
\langle \sigma_{\rm ann} v \rangle=\dfrac{1}{9\,\zeta^2(3)T'^4}\int_{4m_\chi^2}^{\infty}\sigma(s)(s-4m_\chi^2)\,F_{\rm fit}(\sqrt{s}/T')\,ds
\label{eq:chiphiTA}
\end{equation}
where, $F_{\rm fit}(x)=\dfrac{a}{1+b\,e^{c\,x}}$ with $a=0.31,\,b=0.29,\,c=1.03$. 
The numerical and the approximate result are in good agreement with each other: the deviation being $\mathcal{O}(10\% -40\%)$, which is sufficient for our purposes. 
When $T' \ll m_N$, the integral in Eq. \eqref{eq:chiphiTA} is almost constant, thus we have $\langle \sigma_{\rm ann} v \rangle \propto Y_\chi^4/m_N^2$. 
The $1/m_N^2$ scaling of the thermally-averaged cross-section in the low temperature limit is understood from the fact that at energy scales much below $m_N$, one can integrate out the $N$ and generate an effective dimension-five operator, $(\overline{\chi} \chi)\phi^2$. 
Its Wilson coefficient scales as $1/m_N$, hence the cross-section behaves as $1/m_N^2$. 
Using Eq. \eqref{eq:xiFin}, the decoupling temperature is given by
\begin{equation}
    T'_{\rm dec} \sim 10 \left( \frac{10^{-12}}{Y_\nu} \right) \left( \frac{0.01}{Y_\chi} \right)^4 \left(\frac{m_N}{\text{GeV}} \right)^3 \text{ keV} \, .
\end{equation}
It is clearly very sensitive to $Y_\chi$. 
A larger dark sector coupling constant implies that $\chi \chi \leftrightarrow \phi \phi$ will stay in equilibrium for longer: for sufficiently large $Y_\chi$, equilibration persists until $T' < m_\chi$ and the DM $\chi$ does not decouple relativistically. 
Thus, in Fig. \ref{fig:relFO} this bound, displayed in light orange, rules out most of the parameter space for $Y_\chi = 10^{-2}$ but is far less stringent for $Y_\chi = 10^{-4}$.

Another limit is obtained from the condition that the dark sector does indeed thermalise. 
This is necessary in order to write the simple expression for the relic abundance in Eq. \eqref{eq:relicnonrel} and hence relate the DM mass to the seesaw parameters. 
Consider $NN \to \chi \chi$, which has cross-section,
\begin{equation}
    \sigma(NN \to \chi \chi) \simeq \frac{Y_\chi^4}{32\pi s (s-4m_N^2)} \left( 5 \sqrt{s(s-4m_N^2)} + \frac{2m_N^2(5m_N^2 - s)}{2m_N^2 - s } \log\left(\frac{s - 2 m_N^2 + \sqrt{s (s-4 m_N^2 )}}{ s - 2 m_N^2 - \sqrt{s (s-4 m_N^2 )}}\right) \right) \, ,
\end{equation}
again neglecting $m_\phi, m_\chi \ll m_N$. 
In the limit $s\, (\sim T'^2) \ll m_N^2$, the log term is subdominant and the thermally-averaged rate is well approximated by
\begin{align}
    \langle \sigma(NN \to \chi \chi) v \rangle &\simeq \frac{5\,Y_\chi^4}{32\pi m_N^2}\left(\frac{K_1(x')}{K_2(x')}\right)^2 \, ,
\end{align}
where $x' \equiv m_N/T'$. 
Imposing that $n_N \langle \sigma(NN \to \chi \chi) v \rangle=H$, where $H \simeq 1.66 \sqrt{g_*} T^2/M_{Pl}$ is the Hubble rate, corresponds to thermalisation. 
It gives the condition 
\begin{equation}
    x'^{1/2} e^{-x'} \simeq 1.1 \sqrt{g_*} \left(\frac{10^{-3}}{Y_\chi} \right)^4 \left( \frac{10^{-12}}{Y_\nu} \right) \left( \frac{m_N}{10 \text{ GeV}} \right)^2 \, ,
\end{equation}
where we have used Eq. \eqref{eq:xiFin}, as well as the fact that $K_1(x')/K_2(x') \simeq 1$ for $x' \gg 1$, and have assumed a Maxwell-Boltzmann distribution for the $N$. 
If there is no solution to this equation, then the dark sector never thermalises. 
Since the LHS obtains a maximum value of $1/\sqrt{2e}$ at $x' = 1/2$, this puts a bound on the combination $m_N^2/(Y_\nu Y_\chi^4)$. 
For constant $Y_\chi$, the bounds therefore form contours of constant $m_N^2/y_\nu$ (except around $m_N \sim m_W$, since in that case the rate for this process peaks before $\xi$ reaches its maximum value), as shown in Fig. \ref{fig:relFO}. 
They are displayed in light purple in the figure. 
Importantly, the constraint becomes stronger as $Y_\chi$ decreases. 
This places an understandable effective lower bound on $Y_\chi$: since $N, \chi, \phi$ interactions are controlled by $Y_\chi$, it is clear that if this coupling becomes too small, the sector cannot thermalise. 
A very similar constraint can obtained by considering $NN \leftrightarrow \phi \phi$ annihilation.

To summarise the results, the relativistic freeze-out scenario is valid for $10^{-4} \lesssim Y_\chi \lesssim 10^{-2}$, as shown in Fig. \ref{fig:relFO}. 
The two $Y_\chi$-dependent conditions that the DM thermalises and that it decouples relativistically bound $Y_\chi$ from below and above, respectively. 
The allowed sterile neutrino and DM masses increases sharply with $Y_\chi$, from $m_\chi < m_N \lesssim 10$ keV for $Y_\chi = 10^{-4}$ to $m_N \gtrsim 10$ GeV and 1 keV $\lesssim m_\chi \lesssim$ 1 GeV for $Y_\chi = 10^{-2}$.


\section{Sequential freeze-in of the DM}
\label{sec:FI}
There is a second, and qualitatively very different, region of parameter space where the DM abundance is directly linked only to its mass and to the seesaw parameters: the case where $Y_\chi$ is very tiny. 
In this scenario the sterile neutrino is first frozen-in, before freezing in the DM through $N \to \chi \phi$ decays. 
As stated in the previous section, the sterile neutrinos are mainly produced from $W$ and $Z$ boson decays, with yield
\begin{equation}
    Y_N \simeq 2.0 \times 10^{-6} Y_\nu^2 \sum \limits_{V=W,Z} \frac{g_V M_{Pl}}{m_V} \left(1 - \frac{m_N^2}{m_V^2}\right)^2  \left(1+\frac{2m_V^2}{m_N^2}\right) \, .
\end{equation}
The width of the sterile neutrino decay to DM is
\begin{equation}
    \Gamma(N \to \chi \phi) = \frac{Y_\chi^2}{16\pi m_N^3} \sqrt{\lambda(m_N^2, m_\chi^2, m_\phi^2)} \left((m_N + m_\chi)^2 - m_\phi^2 \right) \, ,
    \label{GammaNchiphi}
\end{equation}
where $\lambda$ is the K\"all\'en-Lehmann function. 
The key observation in \cite{Coy:2021sse} is that if the branching ratio of this decay is 1, the DM yield can be found due to the simple relation
\begin{equation}
    Y_\chi = Y_\phi = Y_N \, .
    \label{YscenA}
\end{equation}
Consequently, the relic abundance is
\begin{equation}
    \Omega_{\rm DM} h^2 \simeq 10^{23} Y_\nu^2 \left(\frac{m_\chi + m_\phi}{\text{GeV}} \right) \left( \frac{10 \text{ GeV}}{m_N} \right)^2 \, .
    \label{eq:abundance} 
\end{equation}
As highlighted previously, this depends only on the seesaw parameters, $Y_\nu$ and $m_N$, and the DM mass, where DM is dominantly the heavier of $\chi$ and $\phi$. 
For definiteness, we will always assume it to be $\chi$. 
We can fix $Y_\nu$ as a function of $m_N$ and $m_\chi$ ($\gg m_\phi$) by imposing that the correct relic abundance, $\Omega_{\rm DM} h^2 = 0.12$ \cite{Planck:2018vyg}, is produced: in this way, $Y_\nu \propto m_N/\sqrt{m_\chi}$, neglecting $m_\phi$.
Thus, the lightest neutrino mass becomes
\begin{equation}
    m_{\nu_1} \simeq 3.6 \times 10^{-12} \, \frac{\text{GeV}}{m_\chi} \left( \frac{m_N}{10 \text{ GeV}} \right) \text{ eV}
\end{equation}
The dashed black contours in Fig. \ref{fig:FI} correspond to $m_{\nu 1} = 10^{-6,-12,-18}$ eV. 
As can be seen from this figure, within the allowed parameter space $10^{-12}$ eV $\lesssim m_{\nu 1} \lesssim 10^{-7}$ eV for each value of $Y_\chi$.

\subsection{Constraints}
There are fewer constraints on this scenario than the large $Y_\chi$ case since all new couplings are small and there is no thermalisation. 
Nonetheless, the possibility of a neutrino line, outlined below, is a notable feature of the model.

\begin{figure}[!t]
    \centering
    \includegraphics[width=0.49\textwidth]{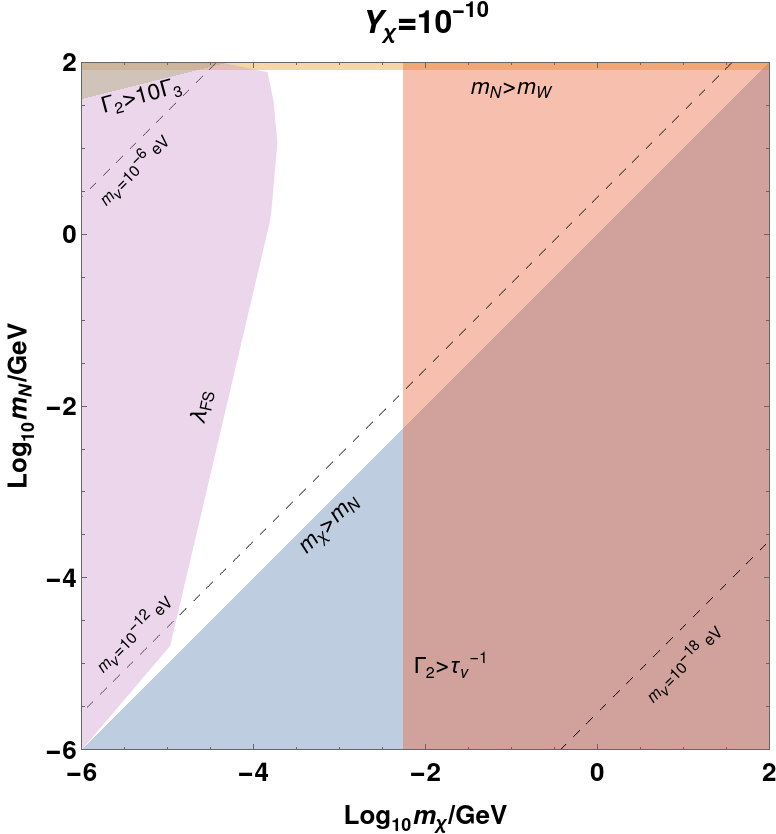}~
    \includegraphics[width=0.49\textwidth]{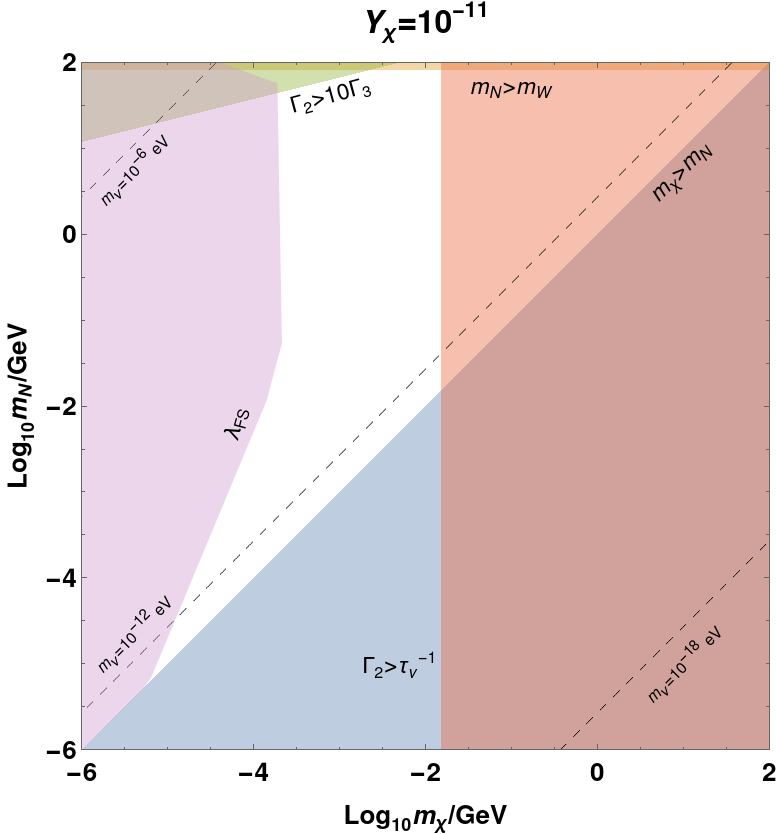}
    \includegraphics[width=0.49\textwidth]{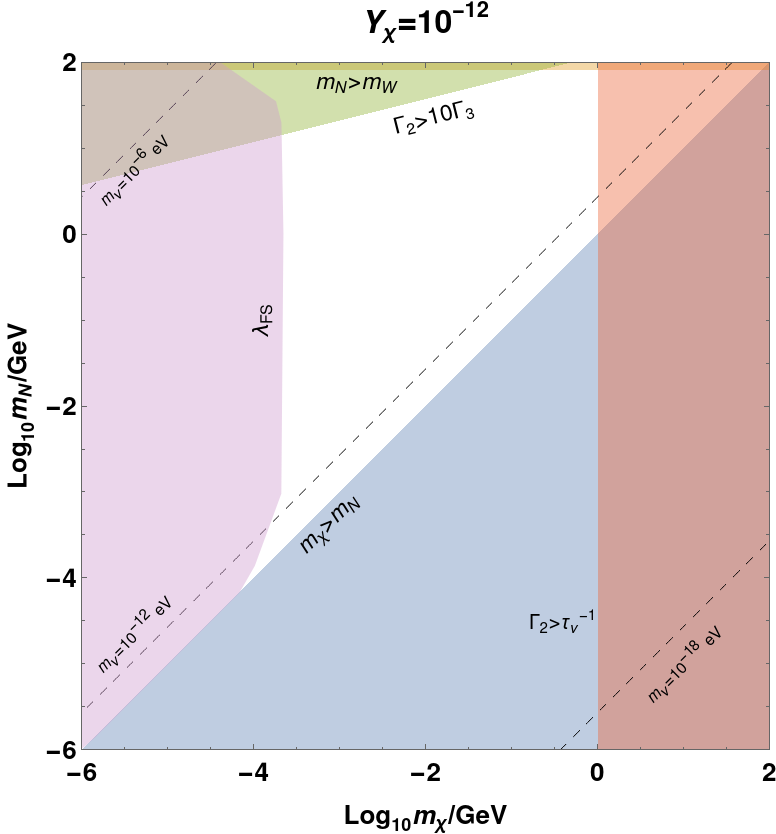}
    \includegraphics[width=0.49\textwidth]{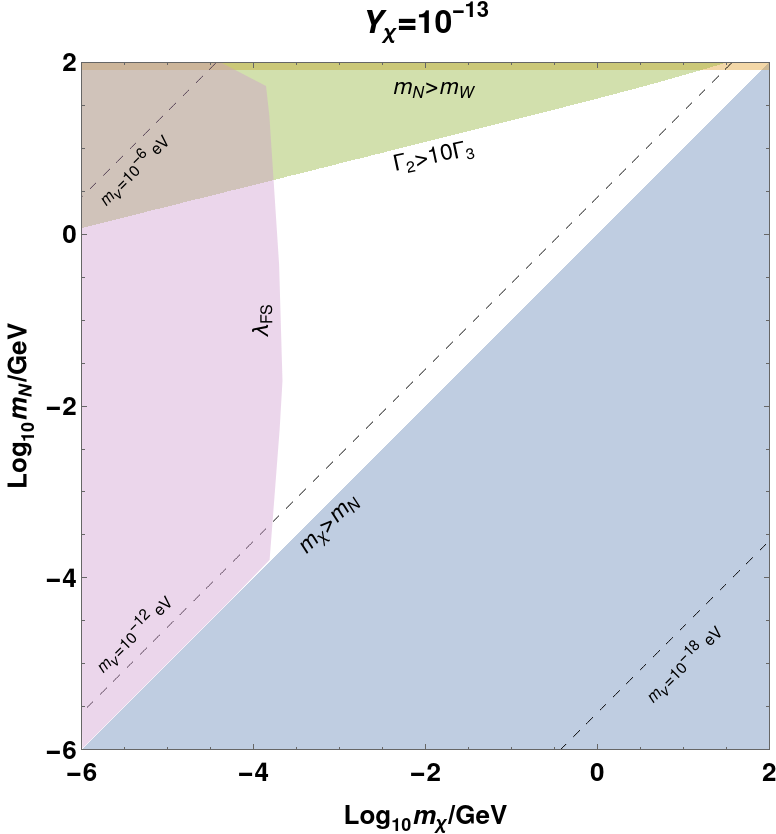}
    \caption{Allowed region for sequential freeze-in scenario for different values of $Y_\chi$. All constraints and contours are indicated in the plots. }
    \label{fig:FI}
\end{figure}

First of all, the simplest constraints are again $m_N < m_W$ (orange, top of each panel of Fig. \ref{fig:FI}) and $m_\chi < m_N$ (blue, bottom right half of each panel), in order for the scenario to be self-consistent, as well as the aforementioned Tremaine-Gunn bound, $m_\chi \gtrsim 1$ keV. 
Moreover, we enforce that the branching ratio $N \to \chi \phi$ is close to 1, so that Eq. \eqref{YscenA} is a sufficiently good approximation. 
Without this, the one-to-one correspondence between DM and seesaw physics would break down. 
In particular, this decay should dominate over three-body decays to SM fermions: $N \to \nu \bar{f} f$ mediated by the $Z$ boson and $N \to \ell \bar{f} f'$ mediated by the $W$ boson. 
The rate of decay to three neutrinos is
\begin{equation}
    \Gamma(N \to \nu \bar{\nu} \nu) = \frac{g_2^2 Y_\nu^2 m_N^3}{2048 \pi^3 c_w^2 m_Z^2} \, , 
\end{equation}
where $c_w$ is the cosine of the weak-mixing angle. 
We sum over all possible three fermion decays, and conservatively assume that all fermions (other than the top quark) are massless. 
The green boundaries in Fig. \ref{fig:FI} are obtained by enforcing that the width of $N \to \chi \phi$ is at least ten times the sum of the three-body decay widths in order to be certain that it dominates. 
Since the bound constrains sterile neutrino masses above GeV, the massless fermion assumption is largely a good approximation. 
Since the three-body decays are independent of $Y_\chi$ while $\Gamma(N \to \chi \phi) \propto Y_\chi^2$, it is clear that the limit becomes stronger with smaller $Y_\chi$. 

Secondly, a weak lower bound can be placed on $Y_\chi$ due to constraints on long-lived particles decaying into radiation. 
It was found in \cite{Hambye:2021moy} that $\tau_\psi f_\psi^2 \lesssim 5 \times 10^9$s, where $\tau_\psi$ is the lifetime of some relic $\psi$, and $f_\psi$ is its fraction of dark matter. 
Assuming that $N \to \chi \phi$ is the dominant sterile neutrino decay, as discussed just above, every $N$ will decay into a single $\chi$ and therefore $f_N = (m_N/m_\chi) f_\chi$. 
If $\chi$ has the correct relic abundance, i.e. $f_\chi = 1$, then we obtain a limit on $\tau_N$, and hence find $Y_\chi \gtrsim 10^{-15} \sqrt{\text{MeV}/m_\chi}$.

One of the most promising bounds comes from the fact that the decay $\chi \to \nu \phi$ could lead to an observable neutrino line. 
The width for this process is \cite{Coy:2021sse}
\begin{equation}
    \Gamma(\chi \to \nu \phi) \simeq \frac{Y_\chi^2 Y_\nu^2}{32\pi} \frac{v^2 m_\chi}{m_N^2} \left( 1 - \frac{m_\phi^2}{m_\chi^2} \right)^2 \simeq 7.2 \times 10^{-24} Y_\chi^2 \text{ GeV} \, ,
\end{equation}
where for the second equality we used Eq. \eqref{eq:abundance} (enforcing that there is the correct relic abundance), and took the limit that $m_\chi \gg m_\phi$. 
The bound depends mainly on $Y_\chi$, with only a mild $m_\chi$-dependence from the fact that the limit on DM two-body decays to a neutrino varies with DM mass, see \cite{Garcia-Cely:2017oco,Coy:2020wxp}. 
The excluded regions are shaded in red in Fig. \ref{fig:FI}. 
As $Y_\chi$ decreases, the DM lifetime increases and hence the constraint vanishes. 
The future detection of such a neutrino line in the keV-GeV range would point towards this class of model with $Y_\chi \sim 10^{-10}-10^{-12}$. 
More generally, CMB data on DM decaying into light species gives $\tau_{\rm DM} > 4.6 \tau_U$ \cite{Poulin:2016nat}, which corresponds to $Y_\chi \lesssim 2 \times 10^{-10}$, further restricting the parameter space.

A more involved constraint comes from structure formation. 
For this, we need the distribution functions of the sterile neutrinos and dark matter, to which we now turn. 

\subsection{Distribution functions of the dark sector particles}
\label{sec:distributions}
In this section, we first compute the distribution function of the sterile neutrino $N$, before using it to compute the $\chi$ and $\phi$ distributions. 
We solve the Boltzmann equation,
\begin{equation}
    L[f_N] = \sum \limits_{N\to ... , ...\to i} \mathcal{C}_N \,,
\end{equation}
where the Liouville operator is 
\begin{equation}
    L[f_N] \equiv \left(\frac{\partial}{\partial t} - H p \frac{\partial}{\partial p}\right) f_N = H x \frac{\partial f_N}{\partial x} \, ,
\end{equation}
where we define $x \equiv m_N/T$ and ignore terms of order $dg_*/dt$, since the number SM relativistic degrees of freedom is generally close to constant. 
On the RHS, we have the collision terms, summing over all processes involving the sterile neutrinos. 
These are computed in appendix \ref{sec:collisions}, and are found to be
\begin{align}
    \mathcal{C}_N (Z \to N \nu) &= \frac{Y_\nu^2 m_Z^2}{8\pi p_N E_N} \left( 1 - \frac{m_N^2}{m_Z^2} \right) \left( 1 + \frac{2m_Z^2}{m_N^2} \right) T (e^{-E_{Z,-}/T} - e^{-E_{Z,+}/T}) \\
    \mathcal{C}_N (W^\pm \to N \ell^\pm) &= \frac{Y_\nu^2 m_W^2}{8\pi p_N E_N} \left( 1 - \frac{m_N^2}{m_W^2} \right) \left( 1 + \frac{2m_W^2}{m_N^2} \right) T (e^{-E_{W,-}/T} - e^{-E_{W,+}/T}) \\
    C_N(N \to \chi \phi) &= - \frac{m_N}{ E_N} \Gamma(N \to \chi \phi) f_N \, ,
\end{align}
where $E_{Z,\pm}$ and $E_{W,\pm}$ are the maximum and minimum kinematically allowed $E_Z$ and $E_W$ and are given in Eqs. \eqref{EZpm} and \eqref{EWpm}. 
Solving the Boltzmann equation therefore gives 
\begin{align}
    H x \frac{\partial f_N}{\partial x} &= \mathcal{C}_N (Z \to N \nu) + \mathcal{C}_N (W^\pm \to N \ell^\pm) - \frac{m_N}{ E_N} \Gamma(N \to \chi \phi) f_N \notag \\
    f_N(x, y_N) &= \exp \left[ \frac{-\Gamma_N}{2x^2H(x)} \left( x \sqrt{x^2 + y_N^2} - y_N^2 \tanh^{-1} \frac{x}{\sqrt{x^2 + y_N^2}} \right) \right] \int_0^{x} dx' \frac{\mathcal{C}_N (Z \to N \nu) + \mathcal{C}_N (W^\pm \to N \ell^\pm)}{x' H(x')} \notag \\
    &\times \exp \left[ \frac{\Gamma_N}{2x'^2 H(x')} \left( x' \sqrt{x'^2 + y_N^2} - y_N^2 \tanh^{-1} \frac{x'}{\sqrt{x'^2 + y_N^2}} \right) \right] \, ,
\end{align}
where $H(x) \equiv 1.66 \sqrt{g_*(x)} m_N^2/(M_{Pl} x^2)$ and where we define $y_N \equiv p_N/T$. 
At late times, $x \to \infty$, the integral becomes constant, $y_N \ll x$, and therefore $f_N \propto \exp[-\Gamma_N/(2H)] = \exp[-\Gamma_N t]$, as should be the case for a decaying particle.

The distribution function of DM produced from decaying thermal particles was precisely computed in \cite{Boulebnane:2017fxw}. 
Our result agrees with their result in the limit $\Gamma \to 0$. 
The distribution function of a particle species produced by decays which then subsequently decays itself does not seem to have been previous computed.

The $\chi$ and $\phi$ are produced via $N$ decay. 
These dark sector distribution functions are the solutions to
\begin{equation}
    H x \frac{\partial f_i}{\partial x} = \mathcal{C}_i(N \to \chi \phi) \, ,
\end{equation}
for $i = \chi, \phi$. 
In appendix \ref{sec:collisions}, we found that the collision terms are
\begin{align}
    C_\chi[N \to \chi \phi] &= \frac{ Y_\chi^2}{16\pi p_\chi E_\chi} \left[ (m_N + m_\chi)^2 - m_\phi^2 \right] \int_{E_{N(\chi),-}}^{E_{N(\chi),+}} dE_N f_N \\
    C_\phi[N \to \chi \phi] &= \frac{ Y_\chi^2}{8\pi p_\phi E_\phi} \left[ (m_N + m_\chi)^2 - m_\phi^2 \right] \int_{E_{N(\phi),-}}^{E_{N(\phi),+}} dE_N f_N \, ,
\end{align}
with integral limits determined by the minimum and maximum kinematically allowed $E_N$, given in Eqs. \eqref{ENChipm} and \eqref{ENPhipm}. 
We can neglect the inverse process since the initial $\chi$ and $\phi$ densities are assumed to be negligible. 
Consequently, we have
\begin{align}
     f_\chi(x_N,y_\chi) &= \frac{M_{Pl}}{1.66 m_N^2} \int_0^{x} dx' \frac{x'}{\sqrt{g_*} } \mathcal{C}_i(N \to \chi \phi) \notag \\
     &= \frac{Y_\chi^2 \left[ (m_N + m_\chi)^2 - m_\phi^2 \right]}{16\pi m_N y_\chi} \int_0^{x} dx' \frac{1}{H(x') \sqrt{y_\chi^2 + x'^2 m_\chi^2/m_N^2}}  \int_{z_{N(\chi),-}}^{z_{N(\chi),+}} dz_N f_N \, ,
\end{align}
where $y_\chi \equiv p_\chi/T$, and similarly
\begin{align}
     f_\phi(x_N,y_\phi) &= \frac{Y_\chi^2 \left[ (m_N + m_\chi)^2 - m_\phi^2 \right]}{8\pi m_N y_\phi} \int_0^{x} dx' \frac{1}{H(x') \sqrt{y_\phi^2 + x'^2 m_\phi^2/m_N^2}}  \int_{z_{N(\phi),-}}^{z_{N(\phi),+}} dz_N f_N \, .
\end{align}
This seems to be the first result for the distribution function of particle species produced via sequential freeze-in.

\begin{figure}
    \centering
    \includegraphics[width=0.7\textwidth]{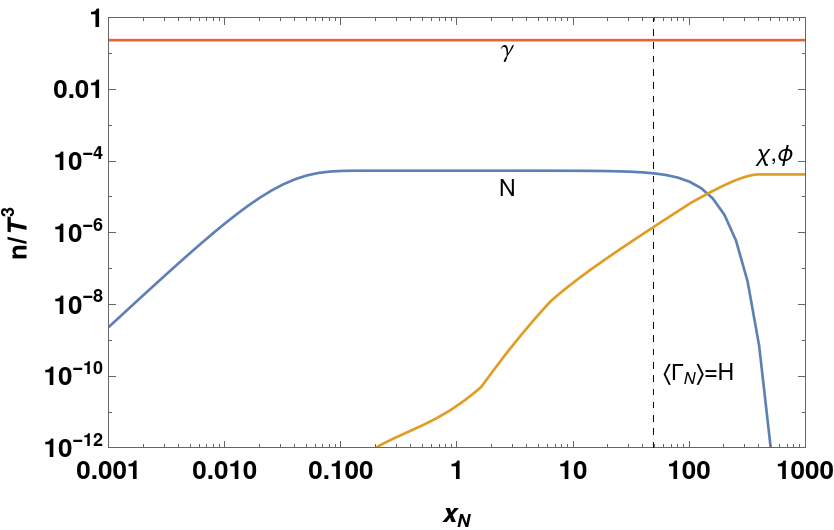}
    \caption{Number densities of the photon (red), sterile neutrino with mass 1 GeV (blue), and $\chi$ and $\phi$ for $m_\chi = 1$ MeV and $m_\phi \ll m_\chi$ (orange). 
    Here $Y_\chi = 10^{-10}$ and $Y_\nu = 3.5 \times 10^{-12}$, with the latter determined by Eq. \eqref{eq:abundance} after fixing $\Omega_{\rm DM} h^2 = 0.12$.
    }
    \label{fig:numbers}
\end{figure}

Having now computed the $\chi$ distribution function, we can calculate its free-streaming horizon and place a bound on the parameter space from structure formation. 
The free-streaming horizon is given by
\begin{equation}
    \lambda_{FS} = \int_{t_i}^{t_0} \frac{\langle v(t) \rangle}{a(t)} dt \, ,
\end{equation}
where $t_i$ is the time of production and $t_0$ is today. 
It is well approximated by
\begin{equation}
    \lambda_{FS} \simeq \frac{\sqrt{t_{eq} t_{nr}}}{a_{eq}} \left( 5 + \ln \frac{t_{eq}}{t_{nr}} \right) \, ,
\end{equation}
where $t_{eq} = 1.9 \times 10^{11}$s is the time of matter-radiation equality, $a_{eq} = 8.3 \times 10^{-5}$ is the corresponding scale factor, and $t_{nr}$ is the time when the DM becomes non-relativistic, i.e. when $\langle p_\chi \rangle = m_\chi$. 
Ly-$\alpha$ constraints on early-decoupled fermionic DM \cite{Irsic:2017ixq} convert into the bound $\lambda_{FS} \lesssim 66$ kpc. 
This is represented by the purple regions in Fig.~\ref{fig:FI}

Finding $\langle p_\chi \rangle = \int d^3p_\chi f_\chi p_\chi/(\int d^3p_\chi f_\chi)$ is computationally challenging, especially at very low temperatures $T \sim m_\chi \ll m_N$. 
We therefore calculated $\langle p_\chi \rangle$ at some time after the $\chi$ had been produced (some $t > \tau_N \equiv \Gamma_N^{-1}$), and used that $p_\chi \propto T$ to solve $\langle p_\chi \rangle = m_\chi$. 
In \cite{Coy:2021sse}, we had assumed that the sterile neutrinos decayed with energy $E_N \simeq m_N$ at $t = \tau_N$, hence $E_\chi = m_N/2$ immediately after the decay. 
Although this captured the correct qualitative behaviour, it in fact leads to bounds about an order of magnitude too strong compared to the more precise limit computed above using the distribution functions.

The $Y_\chi$-dependence of the structure formation bound is most clearly seen in the bottom left of Fig. \ref{fig:FI}, corresponding to light $N$ and $\chi$. 
As $Y_\chi$ decreases, more of this light mass region is ruled out, until $Y_\chi = 10^{-13}$ when all $m_\chi \lesssim 100$ keV is excluded. 
This is understandable as smaller $Y_\chi$ implies later decays of the sterile neutrinos, and hence the DM particles are relatively more energetic at the time of structure formation, $T \sim$ keV.

The computation of the distribution functions also allows us to precisely track the number density of each species over time. 
A simple example is displayed in Fig. \ref{fig:numbers}, taking the case $Y_\chi = 10^{-10}$, $m_N = 1$ GeV, $m_\chi = 1$ MeV and $m_\phi \ll m_\chi$, which falls within the white allowed region of the top-left panel of Fig. \ref{fig:FI}. 
The comoving sterile neutrino number rises until a fixed value: this peak is reached at $T \sim m_{W,Z}/10$ when the $W$ and $Z$ abundances becomes Boltzmann-suppressed and hence their decays are also suppressed. 
As is expected for freeze-in, $n_N \ll n_\gamma$. 
The $\chi$ and $\phi$ number densities slowly rise as the sterile neutrinos start decaying, with the change in the gradient of the orange line in Fig. \ref{fig:numbers} corresponding to when the sterile neutrinos become non-relativistic. 
Soon after $\langle \Gamma_N \rangle = H$, given by the dashed vertical line, the $N$ rapidly decay away to be replaced by the $\chi$ and $\phi$.

Comparing Figs. \ref{fig:relFO} and \ref{fig:FI}, it is clear that the sequential freeze-in case has a much broader allowed parameter space than the relativistic freeze-out case. 
Indeed, the DM-seesaw correspondence holds for $10^{-17} \lesssim Y_\chi \lesssim 10^{-10}$, with the DM and sterile neutrino masses able to be keV, MeV or GeV-scale. 
Again, the lightest neutrino mass eigenstate must be extremely tiny, $m_{\nu 1} \lesssim 10^{-7}$ eV, and this case also has the promising possibility of a detectable neutrino line of energy $m_\chi/2$.

\section{Conclusions}
In this work we have investigated two scenarios of a one-to-one correspondence between dark matter and neutrino physics. After introducing the neutrino portal DM model in section \ref{sec:Model}, we analysed the relativistic freeze-out scenario in detail in section \ref{sec:FO}. 
Here, dark matter thermalised within a dark sector freezes out relativistically and thereby the final relic abundance becomes independent of the neutrino portal strength, being only a function of the seesaw parameters. 
We showed that the portal coupling is bound from below by the requirement that dark sector must thermalise, and from above by the condition that the freeze-out is relativistic. 
By computing thermal averages of the relevant scattering processes we found that the one-to-one correspondence holds (and observational constraints are satisfied) for $10^{-4} \lesssim Y_\chi \lesssim 10^{-2}$. 
Our results for this section are summarised in Fig. \ref{fig:relFO}. 
Notably, the allowed sterile neutrino mass increases sharply with $Y_\chi$: $m_N \sim 1-10$ MeV for $Y_\chi = 10^{-4}$ while $m_N \sim 10-80$ GeV for $Y_\chi \sim 10^{-2}$. The DM mass can vary from keV up to GeV. 
This scenario also predicts that the lightest neutrino should be very light indeed, with $m_{\nu 1} \lesssim 10^{-5}$ eV.

The second way this one-to-one correspondence can occur is when the dark matter is produced via a sequential freeze-in process, considered in section \ref{sec:FI}. 
The allowed parameter of this scenario space shrinks quite slowly with decreasing $Y_\chi$, and we found that $10^{-17} \lesssim Y_\chi \lesssim 10^{-10}$ is permitted. 
The results are summarised in Fig. \ref{fig:FI}. 
Since the parent particle producing the dark matter is itself produced in a non-thermal fashion, we calculated the distribution of the dark matter species from first principles. 
Using this, we constrained the dark matter parameter space from the consideration of limits from structure formation. The improvement is significant compared to a previous, simplistic treatment. 
Unlike in the relativistic freeze-out case, larger sterile neutrino masses open up with smaller $Y_\chi$, with $m_N \lesssim m_W$ allowed even for $Y_\chi \sim 10^{-13}$. 
This sequential freeze-in scenario not only predicts an extremely tiny lightest neutrino mass, $m_{\nu 1} \lesssim 10^{-7}$ eV, it also allows for the possibility of a neutrino line from dark matter decays.

\begin{acknowledgements} 
We thank Anirban Biswas, Quentin Decant, Thomas Hambye, Marco Hufnagel, and Matteo Lucca for helpful discussions. 
This project has received support from the IISN convention 4.4503.15 and the ARC grant. 
\end{acknowledgements}

\appendix
\appendixpage
\addappheadtotoc

\section{Collision terms}
\label{sec:collisions}
To compute the $N$, $\chi$ and $\phi$ distribution functions accurately, as done in section \ref{sec:distributions}, we need to know the collision terms involving these particles. 
The collision term for an initial state particle in the $2 \to 2$ scattering $ab \to ij$ is
\begin{equation}
    \mathcal{C}_a = -\frac{1}{S } \frac{1}{2E_a} \int d\Pi_b \int d \Pi_i \int d\Pi_j \, (2\pi)^4 \delta^{(4)} (p_a + p_b - p_i - p_j) \overline{|\mathcal{M}(ab \to ij)|^2} f_a f_b (1\pm f_i)(1\pm f_j)\, ,
\end{equation}
where $S$ is the symmetry factor which accounts for the multiplicities in the initial and final states, $d\Pi_X = g_X d^3 p_X/((2\pi)^3 2E_X)$ denotes the integration over the phase space of particle $X$ with $g_X$ internal degrees of freedom, and $\overline{|\mathcal{M}(ab\to ij)|^2}$ is the squared matrix element averaged over initial and final state spins.
For a final state particle $i$, we make the replacement $a \leftrightarrow i$ everywhere except in the matrix element, and reverse the sign. For a three-body process there is one fewer momentum integral and one fewer distribution function.

\subsection{$Z \leftrightarrow \nu N$}
Sterile neutrinos are dominantly produced via $Z \to \nu N$ and $W^\pm \to \ell^\pm N$. Consider first the $Z$ decay, which is simpler because we can take $m_\nu = 0$. 
The squared matrix element averaging over spins and polarisations is
\begin{align}
    \overline{|\mathcal{M}(Z \to N \overline{\nu})|^2} = \overline{|\mathcal{M}(Z \to \overline{N} \nu)|^2} = \frac{1}{6} Y_\nu^2 m_Z^2 \left( 1 - \frac{m_N^2}{m_Z^2} \right) \left( 1 + \frac{2m_Z^2}{m_N^2} \right) \, .
\end{align}
Therefore the collision term for $N$, summing over $Z \to N \overline{\nu}$ and $Z \to \overline{N}\nu$, is
\begin{align}
    \mathcal{C}_N (Z \to N \nu) = \frac{Y_\nu^2 m_Z^2}{6E_N} \left( 1 - \frac{m_N^2}{m_Z^2} \right) \left( 1 + \frac{2m_Z^2}{m_N^2} \right) \int d \Pi_Z \int d\Pi_\nu \, (2\pi)^4 \delta^{(4)} (p_Z - p_N - p_\nu) f_Z \, ,
\end{align}
ignoring the $N$ distribution function since the $N$ abundance is initially negligible, and also ignoring the $1 - f_\nu$ Pauli-blocking factor. Since the neutrino Fermi-Dirac distribution, $f_\nu = (e^{E_\nu/T} + 1)^{-1}$, obeys $0 \leq f_\nu \leq 1/2$, the error by neglecting the $1 - f_\nu$ term is at most a factor of 2. 
Taking a Maxwell-Boltzmann distribution for the $Z$, we find
\begin{align}
    \mathcal{C}_N (Z \to N \nu) 
    &= \frac{Y_\nu^2 m_Z^2}{8\pi p_N E_N} \left( 1 - \frac{m_N^2}{m_Z^2} \right) \left( 1 + \frac{2m_Z^2}{m_N^2} \right) T (e^{-E_{Z,-}/T} - e^{-E_{Z,+}/T}) \, ,
\end{align}
where $E_{Z,\pm}$ are the maximum and minimum allowed $E_Z$, given by
\begin{equation}
    E_{Z,\pm } = \frac{(m_Z^2 + m_N^2) E_N \pm (m_Z^2 - m_N^2) p_N}{2 m_N^2} \, .
    \label{EZpm}
\end{equation}
This result agrees with \cite{Boulebnane:2017fxw}.  
The reverse process, $N \nu \to Z$ can be neglected, as is typical in freeze-in, since we assume that the $N$ abundance is small compared to the SM gauge boson abundance.

\subsection{\texorpdfstring{$W^\pm \leftrightarrow \ell^\pm N$}{W -> lN}}
The case of $W^\pm \leftrightarrow \ell^\pm N$ is similar to that of $Z \to N \nu$, except that here we cannot neglect the SM charged fermion mass, since it may be larger than $m_N$. 
The averaged matrix element is
\begin{align}
    \overline{|\mathcal{M}(W^+ \to N \ell^+)|^2} = \overline{|\mathcal{M}(W^- \to N \ell^-)|^2} = \frac{1}{6} Y_\nu^2 m_W^2 \left( 1 - \frac{m_N^2}{m_W^2} \right) \left( 1 + \frac{2m_W^2}{m_N^2} \right) \, ,
\end{align}
hence the collision term summing over $W^\pm$ decays is
\begin{align}
    \mathcal{C}_N (W^\pm \to N \ell^\pm) &= \frac{Y_\nu^2 m_W^2}{8\pi p_N E_N} \left( 1 - \frac{m_N^2}{m_W^2} \right) \left( 1 + \frac{2m_W^2}{m_N^2} \right) T (e^{-E_{W,-}/T} - e^{-E_{W,+}/T}) \, ,
\end{align}
where $E_{W,\mp}$ are the minimum and maximum allowed $E_W$, given by
\begin{equation}
    E_{W,\pm } = \frac{(m_W^2 + m_N^2 - m_l^2) E_N \pm p_N \lambda^{1/2}(m_W^2, m_N^2, m_l^2)}{2m_N^2} \, .
    \label{EWpm}
\end{equation}
Note that $E_{W,\pm} \to E_{Z,\pm}$ in the limit that $m_l \to 0$, as it should. 

\subsection{\texorpdfstring{$N \leftrightarrow \chi \phi$}{N -> chi phi}}
In general, the most efficient interaction between the dark sector particles will be the three-body process $N \leftrightarrow \chi \phi$. 
In particular, in the limit that $Y_\chi \ll 1$, as is the case in the sequential freeze-in discussed in section \ref{sec:FI}, 4-body scatterings and annihilations are highly suppressed due to the extra powers of $Y_\chi$. 
The averaged matrix element of the decay is
\begin{equation}
    \overline{|\mathcal{M}(N \to \chi \phi)|^2} = \frac{1}{2} Y_\chi^2 \left[ (m_N + m_\chi)^2 - m_\phi^2 \right] \, .
\end{equation}
Therefore the collision term for the $N$ is
\begin{align}
    C_N(N \to \chi \phi) &= -\frac{Y_\chi^2}{16\pi p_N E_N} \left[ (m_N + m_\chi)^2 - m_\phi^2 \right] f_N \int_{E_{\chi(N),-}}^{E_{\chi(N),+}} dE_\chi (1 - f_\chi)(1 + f_\phi) \, ,
\end{align}
where the minimum and maximum $\chi$ energies are
\begin{equation}
    E_{\chi(N), \pm} = \frac{(m_N^2 + m_\chi^2 - m_\phi^2 ) E_N \pm p_N \lambda^{1/2}(m_N^2, m_\chi^2, m_\phi^2)}{2m_N^2} \, .
    \label{EChiNpm}
\end{equation}
Taking $f_\chi, f_\phi \ll 1$, and using Eq. \eqref{GammaNchiphi}, we therefore have
\begin{equation}
    C_N(N \to \chi \phi) = - \frac{m_N}{ E_N} \Gamma(N \to \chi \phi) f_N \, .
\end{equation}
The collision term for the $\chi$ can be found similarly,
\begin{align}
    C_\chi[N \to \chi \phi] 
    &= \frac{ Y_\chi^2}{16\pi p_\chi E_\chi} \left[ (m_N + m_\chi)^2 - m_\phi^2 \right] (1- f_\chi) \int_{E_{N(\chi),-}}^{E_{N(\chi),+}} dE_N f_N (1+ f_\phi) \, ,
    \label{eq:Cchi}
\end{align}
with lower limit
\begin{equation}
    E_{N(\chi),\pm} = \frac{( m_N^2 + m_\chi^2 - m_\phi^2) E_\chi \pm p_\chi \lambda^{1/2}(m_N^2, m_\chi^2, m_\phi^2)}{2m_\chi^2} \, .
    \label{ENChipm}
\end{equation}
Finally, for the $\phi$ we have
\begin{align}
    C_\phi[N \to \chi \phi] &= \frac{ Y_\chi^2}{8\pi p_\phi E_\phi} \left[ (m_N + m_\chi)^2 - m_\phi^2 \right] (1 + f_\phi) \int_{E_{N(\phi),-}}^{E_{N(\phi),+}} dE_N f_N (1 - f_\chi)
       \label{eq:Cphi}
\end{align}
There is a factor of 2 enhancement compared to the collision terms for the $N$ and $\chi$ since $g_N g_\chi = 4 = 2g_N g_\phi = 2 g_\chi g_\phi$. 
In this case we write $E_{N,<}(p_\phi)$ as
\begin{equation}
     E_{N(\phi),\pm} = \frac{( m_N^2 + m_\phi^2 - m_\chi^2) E_\phi \pm p_\phi \lambda^{1/2}(m_N^2, m_\chi^2, m_\phi^2)}{2m_\phi^2} \, .
     \label{ENPhipm}
\end{equation}
For both Eq. \eqref{eq:Cchi} and Eq. \eqref{eq:Cphi}, we can neglect $f_\chi, f_\phi \ll 1$, a standard freeze-in approximation.


\bibliography{main.bib}


\end{document}